\documentclass[11pt]{article}
\usepackage{graphicx}
\begin{document}
%

\begin{center}
{\large \bf A Way to Understand the Mass Generation}

\vskip.5cm

W-Y. Pauchy Hwang\footnote{Correspondence Author;
 Email: wyhwang@phys.ntu.edu.tw; an overhauled version of
 arXiv:1301.6464v6 [hep-ph] 28 July 2014.}
 \\
{\em Asia Pacific Organization for Cosmology and Particle Astrophysics, \\
Institute of Astrophysics, Center for Theoretical Sciences,\\
and Department of Physics, National Taiwan University,
     Taipei 106, Taiwan}
\vskip.2cm


{\small(November 3, 2014; Revised: November 3, 2015)}
\end{center}

\begin{abstract}
We believe that the quantum 4-dimensional Minkowski space-time with
the force-fields gauge-group structure $SU_c(3) \times SU_L(2)
\times U(1) \times SU_f(3)$ built-in from the very beginning is the
background for everything. Thus, the self-repulsive, but "related",
complex scalar fields $\Phi(1,2)$ (the Standard-Model Higgs), $\Phi(3,1)$
(the purely family Higgs), and $\Phi(3,2)$ (the mixed family Higgs),
with the first family label and the second $SU_L(2)$ label, co-exist
such that they generate all the masses if, and only if, necessary.
Note that the "ignition" channel is on the elusive $\Phi(3,1)$
channel, yielding the prediction that the Standard-Model (SM)
Higgs mass $m_{SM}$ is half of the SM vacuum expectation value $v$.
Before the "ignition", there is no mass terms, including
the Higgs, the quarks, and the leptons. Apart from the
"ignition" term, all the couplings are dimensionless and
thus the theory is determined by the quantum 4-dimensional
Minkowski space-time. The multi-GeV or
sub-sub-fermi $SU_f(3)$ family gauge fields protect the
lepton world from the QED Landau ghost and make them
asymptotically free. We try to discuss different ways
to deal with infinities (ultraviolet divergences) for
the resultant Standard Model (as the consistent and
complete theory).

\bigskip

{\parindent=0pt PACS Indices: 12.60.-i (Models beyond the standard
model); 98.80.Bp (Origin and formation of the Universe); 12.10.-g
(Unified field theories and models).}
\end{abstract}

\bigskip

\section{Prelude}

The late Professor Henry Primakoff, my Ph.D. advisor, spoke at his
60th Birthday Symposium, beginning with the question, "Why is our
space three-dimensional? Why our time one-dimensional?" In the
late 1970's, I did not know the answer and even did not know this
would be a meaningful question. I bet that Henry knew part of the
answers except that he did not tell the other people.

Until these years (when I reached 60's and when there is no
more pressure for the publications), I recognized that in fact
these are deep meaningful questions. We attempt to describe the
point-like particles by the complex scalar fields, the Dirac
fields, and the force-fields gauge fields - there are a lot of
mysteries (magics) happening with the 4-dimensional Minkowski
space-time. In the 4-dimension, the complex scalar field is
naturally born with the self-repulsive interaction, $\lambda
(\phi^\dagger \phi)^2$ with positive $\lambda$ determined
by the 4-dimensional Minkowski space-time ({\it not} by the
field $\phi$ itself). Thus, the complex scalar fields are
called for as the Higgs fields for the force-fields gauge
fields.

The Dirac fields, such as for the electrons, stems
from the linearization of the Einstein relations,
$E^2={\vec p\,}^2 + m^2$. It calls for the two
components of the spin, in addition to the existence
of the antiparticle. Again, the 4-dimensional Minkowski
space-time is the origin of the spin, and also the
origin of the antiparticle.

Thus, first of all, we stress that we live in the
quantum 4-dimensional Minkowski space-time. In this
4-dimensional Minkowski space-time, a complex scalar
field $\phi(x)$ cannot exist by itself, since the self-repulsive
interaction $\lambda (\phi^\dagger \phi)^2$ is born with it. This
self-repulsive interaction exists because of the 4-dimensional
Minkowski space-time, independent of the field $\phi(x)$ itself.
This is a magic for the 4-dimensional Minkowski space-time.
$\lambda$ is dimensionless (and equals to ${1\over 8}$) in this
4-dimensional Minkowski space-time.

So, we live in the (quantum) 4-dimensional Minkowski space-time
with the force-fields group $SU_c(3) \times SU_L(2) \times U(1)
\times SU_f(3)$ gauge-group structure built-in from the very
beginning \cite{Fields}. The complex scalar (Higgs) fields now exist in
pairs: $\Phi(1,2)$ (SM Higgs), $\Phi(3,2)$ (mixed family Higgs),
and $\Phi(3,1)$ (purely family Higgs) jointly to make weak bosons $W^\pm$,
$Z^0$ and the family bosons massive. This is the "background" of our
world.

The quark world and, separately, the lepton world are
accepted by the "background" of our world \cite{Fields}.
They all come from Dirac's linearization of the Einstein relation
$E^2={\vec p\,}^2+m^2$. The size of the quark world is typically
$(1\, fermi)^3$, while that of the lepton world of the atom
size ($\sim (10^{-8}cm)^3$).

This is basically the rationale of "our world" - the building
blocks of everything. Those things exist and can be felt by us,
and the rest does not exist.

Thus, we are eventually coming to the following global picture \cite{Fields}:
Our world is basically the quantum 4-dimensional Minkowski
space-time with the force-fields gauge-group structure
$SU_c(3) \times SU_L(2) \times U(1) \times SU_f(3)$ built-in
from the outset. This offers the "background" which accepts
the quark world in view of its (123) symmetry and also
accepts the lepton world of another (123) symmetry.
The quark world has the characteristics with the length
scale so much different from those of the lepton world -
so, the acceptance story by the "background" is also
different.

We emphasize that the quark world is accepted by the
"background" in view of its (123) symmetry (i.e., under
$SU_c(3) \times SU_L(2) \times U(1)$). It is well-behaved
when the energy is very large, or $Q^2\to \infty$, or
when the distance between two quarks is very small, or
$r\to 0$. We propose that the lepton world is protected
by another (123) symmetry (i.e., under $SU_L(2) \times
U(1) \times SU_f(3)$) - so, well-behaved as $Q^2 \to
\infty$, or $r\to 0$.

In fact, we try to regard "the quark world" or "the lepton
world" not just as "a physical system" but also as "a
mathematical system".

In the very beginning, we have only the "background" - the
quantum 4-dimensional Minkowski space-time with the force-fields
$SU_c(3) \times SU_L(2) \times U(1) SU_f(3)$ gauge-group
structure built-in from the outset. The complex scalar
fields $\Phi(1,2)$ (Standard-Model Higgs), $\Phi(3,1)$
(purely family Higgs), and $\Phi(3,2)$ (mixed family Higgs)
enter to make the gauge bosons massive; all other complex scalar
fields, due to lack of "relatives", are self-repulsive
and thus {\it do not exist}.

The matter exists in the linear form of the basic Einstein
relation $E^2={\vec p\,}^2 + m^2$. Thus, the solutions of
the Dirac equations become the only acceptable forms of matter.
This apply to every form of matter, including the leptons
(such as electrons and neutrinos) and the quarks.

To be more specific, in the 4-dimensional Minkowski
space-time, what is the behavior of a complex scalar
field $\phi(x)$? A complex scalar field would have
the following Lagrangian:
\begin{equation}
L = - (\partial_\mu\phi)^\dagger\partial_\mu \phi - \{m^2 \phi^\dagger \phi
+ \lambda (\phi^\dagger \phi)^2\},
\end{equation}
with the last two terms defining the minus of the potential, $-V(x)$.

$\lambda$ has to be positive to stabilize the system.
In only the 4-dimensional Minkowski space-time, $\lambda$
is dimensionless - it must be
a universal constant independent of the field itself,
or, it is a characteristic of the 4-dimensional
Minkowski space-time. We suspect that $\lambda =
{1\over 8}$, although at present we do not know
how to prove this basic aspect.

A positive $\lambda$ means that this self-interaction
is repulsive, and so it cannot exist by itself. Note that,
when the temperature is high enough (like in the early
Universe), the mass term becomes irrelevant (i.e., very
small by comparison).

Thus, the lagrangian for the single complex scalar field $\phi$
is fixed (is given) except that the mass term could be
adjusted, "fixed" in the 4-dimensional Minkowski space-time. The
complex scalar field $\phi$ is self-repulsive and cannot exist.

That is why we must introduce the related complex
scalar fields $\Phi(3,2)$ (mixed family Higgs) and
$\Phi(3,1)$ (purely family Higgs) to lower the
energy and to make the entire story. (In the
notations, the first number refers to $SU_f(3)$
while the second for $SU_L(2)$.) This is the
story on the origin of mass \cite{Origin}.

It is just right to have the three Higgs fields,
and only the three, - $\Phi(1,2)$ (the
Standard-Model Higgs), $\Phi(3,2)$ (mixed family
Higgs), and $\Phi(3,1)$ (purely family Higgs) -
to make the family gauge bosons all massive and
to make the proper room for neutrino oscillations.

Our world is the (quantum) 4-dimensional Minkowski space-time
and the structure, the group structure, is the $SU_c(3) \times
SU_L(2) \times U(1) \times SU_f(3)$. It means that every
object has the designated group property and it transforms in
a certain way under the 4-dimensional Lorentz group. As the nature
is described by the Standard Model, every object should have the
designated group and Lorentz group transformation properties.

The three "related" Higgs, being the complex scalar fields, act
as the systems of energies, self-interacting (and dimensionless)
via $\lambda (\phi^\dagger \phi)^2$ and interacting
equivalently with other Higgs. We conclude \cite{Origin}
that these related three Higgs interact attractively with
a universal $\lambda$. When the temperature is low enough,
it becomes the "mass" phase, or the phase in which
the particles have masses.

\bigskip

\section{The Quark World}

In the Standard Model \cite{Hwang417}, we live in the quantum
4-dimensional Minkowski space-time with the force-field
gauge-group structure $SU_c(3) \times SU_L(2) \times U(1)
\times SU_f(3)$ built-in from the outset. This is what we call
"the background".

Then, what is the quark world? The quark world is a world of matter
form, thus of the type of Dirac equations. It claims a rather small
length scale, of about $10^{-13}cm$. The strong-interaction
nature of $SU_c(3)$ explains such small size. The color $SU_c(3)$
gauge fields are already classified as part of "the background" -
the quarks of three colors and of six flavors are building blocks
of matter for the quark world. The quark world knows the gauge
group $SU_c(3) \times SU_L(2) \times U(1)$, but not $SU_f(3)$ --
the so-called (123) symmetry.

The quark world knows color $SU_c(3)$ well - the strong interaction
that acts in the range of fermi's (i.e., $10^{-13}\, cm$). Everything
larger than a few fermi would eventually cut off the influence of
the strong interaction, unless some special arrangements are given
(by the God).

The lepton world is very similar except that the scale is much
bigger, at the atomic scale, or $10^{-8}\, cm$. But it does not
know the color $SU_c(3)$, except indirectly.

"Our world" is the combination of the background, the quark
world, and the lepton world - so, it is quite complicated
but in fact all of them are (quantum) point-like particles.
Amazingly, they could be represented as a branch of
mathematics - or, relativistic quantum mechanics and
quantum field theory \cite{Book}.

The decomposition of the Standard Model could make our thoughts
much clearer, eventually to adopt a language which is precise
\cite{definition}. Such as: we live in the (quantum) 4-dimensional
Minkowsi space-time with the force-fields gauge-group structure
$SU_c(3) \times SU_L(2) \times U(1) \times SU_f(3)$ built-in
from the outset - as the "background" of our world.
This background supports the quark world. For some reason, it also
supports the lepton world.

In introducing the family concept as a gauge group, we
regard \cite{HwangYan} $((\nu_\tau,\,\tau)_L,\,(\nu_\mu,\,\mu)_L,
\,(\nu_e,\,e)_L)$ $(columns)$ ($\equiv \Psi(3,2)$) as the
$SU_f(3)$ triplet and $SU_L(2)$ doublet.
It is essential to complete the (extended) Standard Model \cite{Hwang417}
by working out the Higgs dynamics in detail \cite{Origin}. It is also
essential to realize the role of neutrino oscillations - it is the change
of a neutrino in one generation (flavor) into that in another generation;
or, we need to have the coupling $i h \bar \Psi_L(3,2)\times
\Psi_R(3,1) \cdot \Phi(3,2)$, exactly the coupling introduced by Hwang
and Yan \cite{HwangYan}. Then, it is clear \cite{Hwang417} that the
mixed family Higgs $\Phi(3,2)$ must be there. The remaining purely
family Higgs $\Phi(3,1)$ helps to complete the picture, so that
the eight gauge bosons are massive in the $SU_f(3)$ family gauge
theory \cite{Family}.

Thus, we see $SU_f(3)$ in the lepton world but it seems that
the quark world does not see $SU_f(3)$ at all. Maybe in the
quark world the $SU_f(3)$ forces are much too feeble than the
$SU_c(3)$ forces.

Another point might be critical. In the quark world, we cannot get
a hand on the mass of an individual quark, because we cannot see
an isolated quark. So, the issue might be how, for us, to reach
the meaning of "mass" for a composite system, such as a
three-quark system, the quark-antiquark system, etc.

Remember that the story is pretty much fixed if the so-called
"gauge-invariant derivative", i.e. $D_\mu$ in the kinetic-energy
term $-\bar \Psi \gamma_\mu D_\mu \Psi$, is given for a given
basic unit \cite{Book}. It seems that this aspect is as fundamental
as the Einstein relation, $E^2={\vec p}^2+m^2$.

Thus, we have, for the up-type right-handed quarks $u_R$, $c_R$,
and $t_R$,
\begin{equation}
D_\mu = \partial_\mu - i g_c {\lambda^a\over 2} G_\mu^a -
i {2\over 3} g'B_\mu,
\end{equation}
and, for the rotated down-type right-handed quarks $d'_R$, $s'_R$,
and $b'_R$,
\begin{equation}
D_\mu = \partial_\mu - i g_c {\lambda^a\over 2} G_\mu^a -
i (-{1\over 3}) g' B_\mu.
\end{equation}

On the other hand, we have, for the $SU_L(2)$ quark doublets such as
$(u_L,\,d'_L)$,
\begin{equation}
D_\mu = \partial_\mu - i g_c {\lambda^a\over 2} G_\mu^a - i g
{\vec \tau\over 2}\cdot \vec A_\mu - i {1\over 6} g'B_\mu.
\end{equation}

That is, we are using $d'_R$, $d'_L$, etc., consistently. In the
quark world, the down-type quarks are always rotated - that means
that the so-called GIM are always there.

The mass term from the old Standard-Model way is given by
\begin{eqnarray}
L_m&=& -G_1 \{ {\bar d}'_R \Phi^\dagger(1,2) Q_{1,L} + h.c. \}
       -G'_1 \{ {\bar s}'_R {\tilde \Phi}^\dagger(1,2) Q_{1,L} +h.c. \}\nonumber\\
&&   - G_2 \{ {\bar s}'_R \Phi^\dagger(1,2) Q_{2,L} + h.c.\}
     - G'_2 \{ {\bar c}_R {\tilde \Phi}^\dagger(1,2) Q_{2,L} + h.c. \}\nonumber\\
&&   - G_3 \{ {\bar b}'_R \Phi^\dagger(1,2) Q_{3,L} +h.c.\}
     - G'_3 \{ {\bar t}_R {\tilde \Phi}^\dagger(1,2) Q_(3,L) + h.c. \}.
\end{eqnarray}
Note that the six couplings $G_{1,2,3}$ and $G'_{1,2,3}$
in principle can be adjusted. The unitary mixings, the so-called
GIM mechanism, for the down-type quarks helps to forbid the
weak neutral current, at the expense of introducing
peculiar cross-mass terms. How to detect these "peculiar"
interactions through the SM Higgs studies would be
something of importance and urgency.

Assuming that the $u$ and $d$ be massless ($G_1\sim G'_1 \sim 0$),
we have to assume four parameters (couplings) to generate four
masses. It seems possible to take $m(t)/m(b) = m(c)/m(s)$, so
then there is some (hidden) symmetry.

In general, the rotation among the down-type quarks means that
there are off-diagonal masses such as $m_{ds}$, etc. In fact,
this is something which we cannot avoid, because $G_1$, $G_2$,
and $G_3$ are three different mass parameters.

In any event, we so far cannot say too much for the "individual"
quark masses for the quark world. It seems that the GIM mechanism,
while cutting of the neutral weak currents, introduces the
off-diagonal mass terms, in a more hidden part of the story.

\bigskip

\section{The Lepton World}

We could follow the following logic in reaching the
global picture. Neutrino oscillations tell us that there
exists an interaction, such as $i {\bar \Psi}_L \times
\Psi_R \cdot \Phi$, where the scalar field and the
left-handed and right-handed fermions are $SU_f(3)$
triplets. We know that the left-handed fermion has to
be the doublet under $SU_L(2)$ - that pushes the
scalar field $\Phi$ also an doublet under $SU_L(2)$.
Thus, we have an interaction of the form,
$i {\bar \Psi}_L(3,2) \times \Psi_R(3,1) \cdot
\Phi(3,2) + h.c.$.

Since we put all six objects as a representation
in the group, we agree that all these objects are
point-like Dirac particles - {\it not} the mixture
of Dirac particles and Majorana particles.
Moreover, particles of the second or third
generation must be of the same characteristics
as the particles of the first generation. All these
are the group theory in mathematics - we physicists
sometime forget the mathematics ABC.

Of course, we are too far in proving experimentally
that these neutrinos are also point-like Dirac
particles. The regularities, or the symmetries, sort
of give us the confidence in all this regarding the
Standard Model.

In the lepton world, we introduce the family triplet,
$(\nu_\tau^R,\,\nu_\mu^R,\,,\nu_e^R)$ (column), under $SU_f(3)$.
Since the minimal Standard Model does not see the right-handed
neutrinos, it would be a natural way to make an extension of the
minimal Standard Model. Or, we have, for $(\nu_\tau^R,\,
\nu_\mu^R,\,\nu_e^R)$,
\begin{equation}
D_\mu = \partial_\mu - i \kappa {\bar\lambda^a\over 2} F_\mu^a.
\end{equation}
and, for the left-handed $SU_f(3)$-triplet and $SU_L(2)$-doublet
$((\nu_\tau^L,\,\tau^L),\, (\nu_\mu^L,\,\mu^L),\, (\nu_e^L,\,e^L))$
(all columns),
\begin{equation}
D_\mu = \partial_\mu - i \kappa {\bar\lambda^a\over 2} F_\mu^a - i g
{\vec \tau\over 2} \cdot \vec A_\mu + i {1\over 2} g' B_\mu.
\end{equation}
The right-handed charged leptons form the triplet $\Psi_R^C(3,1)$ under
$SU_f(3)$, since it were singlets their common factor $\bar\Psi_L(\bar 3,2)
\Psi_R(1,1)\Phi(3,2)$ for the mass terms would involve the cross terms such as
$\mu\to e$.

The neutrino mass term assumes the {\it unique} form:
\begin{equation}
i {h\over 2} {\bar\Psi}_L(3,2) \times \Psi_R(3,1) \cdot \Phi(3,2)
+ h.c.,
\end{equation}
Here the Higgs field $\Phi(3,2)$ is the mixed family Higgs, because
it carries some nontrivial $SU_L(2)$ charge. In fact, the charged
part of $\Phi(3,2)$ does not experience the spontaneous
symmetry breaking (SSB), as worked out explicitly in \cite{Origin}.

We wish to note, again, that, for charged leptons, the Standard-Model
choice is $\Psi^\dagger(\bar 3,2) \Psi_R^C(3,1) \Phi(1,2) +c.c.$, which
gives three leptons an equal mass. But, in view of that if
$(\phi_1,\phi_2)$ is an $SU(2)$ doublet then $(\phi_2^\dagger,
-\phi_1^\dagger)$ is another doublet, we could form
${\tilde\Phi}^\dagger(3,2)$ from the doublet-triplet $\Phi(3,2)$.

\begin{equation}
i {h^C\over 2} {\bar\Psi}_L(3,2) \times \Psi_R^C(3,1) \cdot
{\tilde \Phi}^\dagger(3,2) + h.c.,
\end{equation}
which gives rise to the imaginary off-diagonal (hermitian) elements
in the $3\times 3$ mass matrix, so removing the equal masses of the
charged leptons.

In the quark world, the GIM mechanism via the down-type quark
mixings, such as Eq. (5) in the previous section, seems to
explain the quark masses - with the six couplings $G_{1,2,3}$
and $G'_{1,2,3}$. In contrast, the lepton world has only
two couplings $h$ and $h^C$, controlling the neutrinos
and the charged leptons - $h$ would be very small in
light of the tiny neutrino masses, and $h^C$ about the $\tau$
or $\mu$ mass. For both the quark world and the lepton
world, these are dimensionless couplings; the same for
the force-fields $SU_c(3) \times SU_L(2) \times U(1) \times
SU_f(3)$ built-in from the outset gauge-group structure.

Because everything is dimensionless, it has to do with
the 4-dimensional Minkowski space-time; it has nothing
to do with the individual fields. Hence, this story is
absolutely beautiful - the act of the Einstein relation
and of its Dirac's linearization in the quantum
4-dimensional Minkowski space-time.

\bigskip

\section{Uniqueness of the Background}

In our publication on "The Origin of Mass" \cite{Origin}, we realize that,
before the spontaneous symmetry breaking (SSB), the Standard Model
does not contain any parameter that is pertaining to "mass", but, after the SSB,
all particles in the Standard Model acquire the mass terms as it should - a way
to explain "the origin of mass". In this way, we sort of tie "the
origin of mass" to the effects of the SSB, or the generalized Higgs mechanism.

That sets the unique stage for the dimensionless interaction $\lambda
(\phi^\dagger \phi)^2$ in the 4-dimensional Minkowski space-time. And
we have the three Higgs fields and the (elusive) purely family Higgs
$\Phi(3,1)$ may work as the "ignition" channel.

Thus, we have to have the various Higgs at our disposal, but not too
many in view of "minimum Higgs hypothesis" or the repulsive nature of
these scalar fields. In the model \cite{Hwang417},
we have the Standard-Model Higgs $\Phi(1,2)$, the purely family Higgs
$\Phi(3,1)$, and the mixed family Higgs $\Phi(3,2)$, with the first label
for $SU_f(3)$ and the second for $SU_L(2)$. We need another triplet $\Phi(3,1)$
since all eight family gauge bosons are massive \cite{Family}.

In another arXiv paper \cite{definition}, we try to introduce the joint-group
space, "$SU_c(3) \times SU_L(2) \times U(1) \times SU_f(3)$ Minkowski
space-time", in the effort of trying to find out what would be the
constraints on the complex scalar fields. First of all, we have to recognize
the special importance of the dimensionless interaction $\lambda
(\phi^\dagger \phi)^2$, the only pure number $\lambda$ for the
4-dimensional low-spin fields. We find $\lambda={1\over 8}$, without
knowing the underlying reason. Secondly, those unrelated complex
fields could be described by $\lambda (\phi_a^\dagger \phi_a
+ \phi_b^\dagger \phi_b)^2$ (with $a \not= b$), through a repulsive
interaction. Thus, we can write an "attractive" interaction,
$(\phi_a^\dagger\cdot \phi_b) \cdot (\phi_b^\dagger\cdot \phi_a)$,
for only those related complex fields. We use this to understand
the origin of mass \cite{Origin}.

Let us write down the terms for potentials among the three Higgs fields,
subject to (1) that they are renormalizable, and (2) that symmetries
are only broken spontaneously (the Higgs or induced Higgs mechanism).
We write \cite{Book,Hwang417}

\begin{eqnarray}
V = & V_{SM} +  V_1 + V_2 + V_3,\\
V_{SM} =& \mu^2 \Phi^\dagger(1,2) \Phi(1,2) + \lambda
(\Phi^\dagger(1,2) \Phi(1,2))^2\\
V_1 =& M^2 \Phi^\dagger(\bar 3,2) \Phi(3,2) + \lambda_1
(\Phi^\dagger(\bar 3,2) \Phi(3,2))^2\nonumber\\
  &+  \epsilon_1(\Phi^\dagger(\bar 3,2)\Phi(3,2))(\Phi^\dagger(1,2)\Phi(1,2))
   + \eta_1 (\Phi^\dagger(\bar 3,2)\Phi(1,2))(\Phi^\dagger(1,2)\Phi(3,2))
  \nonumber\\
  &+ \epsilon_2(\Phi^\dagger(\bar 3,2)\Phi(3,2))(\Phi^\dagger(\bar 3,1)\Phi(3,1))
   + \eta_2 (\Phi^\dagger(\bar 3,2)\Phi(3,1))(\Phi^\dagger(\bar 3,1)\Phi(3,2))
   \nonumber\\
  & + (\delta_1 i \Phi^\dagger(3,2)\times \Phi(3,2) \cdot \Phi^\dagger(3,1) +h.c.),\\
V_2 =& \mu_2^2 \Phi^\dagger(\bar 3,1)\Phi(3,1) + \lambda_2
(\Phi^\dagger(\bar 3,1) \Phi(3,1))^2 \nonumber\\
& + (\delta_2 i \Phi^\dagger(3,1)\cdot
\Phi(3,1) \times \Phi^\dagger(3,1) +h.c.)\nonumber\\
 &+ \lambda_2^\prime\Phi^\dagger(\bar 3,1) \Phi(3,1) \Phi^\dagger(1,2) \Phi(1,2),\\
V_3 =& (\delta_3 i \Phi^\dagger(3,2)\cdot \Phi(3,2)\times
    (\Phi^\dagger(1,2)\Phi(3,2)) +h.c.)\nonumber\\
    &+ (\delta_4 i (\Phi^\dagger(3,2)\Phi(1,2))\cdot \Phi^\dagger(3,1)
    \times \Phi(3,1)+h.c.)\nonumber\\
    & + \eta_3 (\Phi^\dagger(\bar 3,2)\Phi(1,2)\Phi(3,1)+c.c.).
\end{eqnarray}

In the U-gauge, we choose to have
\begin{equation}
\Phi(1,2)= (0,{1\over \sqrt 2} (v+\eta)),\,\, \Phi^0(3,2) = {1\over \sqrt 2} (u_1+\eta'_1, u_2+
\eta'_2, u_3+\eta'_3 ),\,\, \Phi(3,1) = {1\over \sqrt 2}(w+\eta',0,0),
\end{equation}
all in columns. The five components of the complex triplet $\Phi(3,1)$ get
absorbed by the $SU_f(3)$ family gauge bosons and the neutral part of
$\Phi(3,2)$ has three real parts left - together making all eight family
gauge bosons massive.

Before the mixing, the masses of the various Higgs are given by
\begin{eqnarray}
\eta:& (\mu^2/\lambda) + {1\over 4}(\epsilon_1+ \eta_1) u_i u_i + {\lambda'_2\over 4}
w^2, \nonumber\\
\eta':& (\mu_2^2/\lambda_2) + {\epsilon_2\over 4} u_i u_i +{\eta_2\over 4} u_1^2 +
{\lambda'_2\over 4} v^2,
\nonumber\\
\eta'_1:& M^2 + {1\over 4}(\epsilon_1+ \eta_1) v^2 +{1\over 4}(\epsilon_2+\eta_2)w^2 +
(\lambda_1-term),\nonumber\\
\eta'_{2,3}:& M^2 + {1\over 4}(\epsilon_1+ \eta_1) v^2 + {\epsilon_2\over 4} w^2
+ (\lambda_1-term),\nonumber\\
\phi_1:& M^2 + {1\over 2}\epsilon_1 v^2 + {1\over 2}\epsilon_2 w^2 + {1\over 2}\eta_2 w^2 +
{\lambda_1\over 2} u_i u_i, \nonumber\\
\phi_{2,3}:& M^2 + {1\over 2}\epsilon_1 v^2 + {1\over 2} \epsilon_2 w^2
+ {\lambda_1 \over 2} u_i u_i.
\end{eqnarray}
The mixing term looks like, apart from some common factor:
\begin{equation}
2 (\epsilon_1+\eta_1)u_i\eta'_i v \eta + 2\epsilon_2 u_i\eta'_i w \eta' +
2 \eta_2 u_1\eta'_1 w\eta' + 2\lambda'_2 w\eta' v\eta.
\end{equation}

For the sake of simplicity, we will neglect the mixing (and the mixing inside
$\eta'_{1,2,3}$) in this paper. To work out on "the origin of mass", we would
drop out all "mass" terms to begin with.

In treating the problem with the renormalization group (RG) equations,
we realize that, even though to begin withwe set all the mass
terms to zero, they would climb back so easily in the case of
the complex scalar fields - as judged by the RG flow diagrams.
This is why have to analyze different problems from a general
lagrangian as in \cite{Hwang417}.

Basically, we can write down a general renormalizable
lagrangian. But ....  Because of three "cooperative"
complex scalar Higgs fields, because of a universal
$\lambda$, and because of only one "ignition" point,
the real lagrangian becomes rather simple \cite{Origin}.
It seems that $lambda$ ($={1\over 8}$) does not subject
to renormalization, owing to that it is determined by
the 4-dimensional Minkowski space-time, {\it not}
by the complex scalar fields themselves.

Let us illustrate typical results of \cite{Origin}.
We begin with \cite{Origin}

\begin{eqnarray}
V_{Higgs} =& \mu^2_2 \Phi^\dagger(3,1) \Phi(3,1) + \lambda
(\Phi^\dagger(1,2) \Phi(1,2)+ cos\theta_P\Phi^\dagger(3,2)\Phi(3,2))^2\nonumber\\
    &  + \lambda(-4 cos\theta_P)
(\Phi^\dagger(\bar 3,2)\Phi(1,2))(\Phi^\dagger(1,2)\Phi(3,2))
  \nonumber\\
  &+\lambda
(\Phi^\dagger(3,1) \Phi(3,1)+ sin\theta_P \Phi^\dagger(3,2)\Phi(3,2))^2
    + \lambda(-4 sin\theta_P)
(\Phi^\dagger(\bar 3,2)\Phi(3,1))(\Phi^\dagger(3,1)\Phi(3,2))
  \nonumber\\
     & + \lambda'_2 \Phi^\dagger(\bar 3,1)\Phi(3,1) \Phi^\dagger(1,2) \Phi(1,2)
  + (terms\,\, in\,\, i\delta's\,\, and\,\, in\,\, decay).
\end{eqnarray}
These are two prefect squares minus the other extremes, to guarantee
the positive definiteness, when the minus $\mu^2_2$ was left out.
($\theta_P$ may be referred to as "Pauchy's angle".)

From the expressions of $u_iu_i$ and $v^2$, we obtain
\begin{equation}
v^2 (3 cos^2\theta_P-1) = sin\theta_P cos\theta_P w^2.
\end{equation}
And the SSB-driven $\eta'$ yields
\begin{equation}
w^2 (1-2 sin^2\theta_P) = - {\mu_2^2\over \lambda} +
(sin2\theta_P - tan\theta_P) v^2.
\end{equation}
These two equations show that it is necessary to have the driving
term, since $\mu^2_2=0$ implies that everything is zero. Also,
$\theta=45^\circ$ is the (lower) limit.

The mass squared of the SM Higgs $\eta$ is $2\lambda
cos\theta_P u_i u_i$ (noting the factor of two), as known to
be $(125\,\, GeV)^2$. The famous $v^2$ is the number
divided by $2\lambda$, or $(125\,\,GeV)^2/(2\lambda$). Using PDG's for
$e$, $sin^2\theta_W$, and the $W$-mass \cite{PDG}, we find
$v^2=255\,\, GeV$. So, we set $\lambda={1\over 8}$, a simple model indeed.

The mass squared of $\eta'$ is $-2(\mu_2^2-sin\theta_P u_1^2 +
sin\theta_P (u_2^2+u_3^2))$. The  other condensates are $u_1^2= cos\theta_P v^2
+ sin\theta_P w^2$ and $u_{2,3}^2 = cos\theta_P v^2 - sin\theta_P w^2$ while
the mass squared of $\eta'_1$ is $u_1^2\lambda$, those of $\eta'_{2,3}$ be
$u_{2,3}^2\lambda$. The mixings among $\eta'_i$ themselves are neglected
in the paper.

There is no SSB for the charged Higgs $\Phi^+(3,2)$. The mass
squared of $\phi_1$ is $\lambda(cos\theta_P v^2 - sin\theta_P w^2)
+ {\lambda\over 2} u_i u_i$ while $\phi_{2,3}$ be $\lambda(cos\theta_P
v^2 + sin\theta_P w^2) + {\lambda\over 2} u_i u_i$.

A further look of these equations tells that $3cos^2\theta_P - 1 > 0$ and
$2sin^2\theta_P -1 > 0$. A narrow range of $\theta_P$ is allowed (greater
than $45^\circ$ while less than $57.4^\circ$, which is determined by
the group structure). For illustration, let us choose
$cos \theta_0 = 0.6$ and work out the numbers as follows:
(Note that $\lambda={1\over 8}$ is used.)
\begin{eqnarray}
& 6 w^2 = v^2, \quad -\mu^2_2/\lambda = 0.32 v^2;\nonumber\\
\eta: & m^2(\eta) =(125\, GeV)^2, \quad v^2 = (250\,GeV)^2;
\nonumber\\
\eta': & m^2(\eta') = (51.03\,GeV)^2, \quad w^2=v^2/6; \nonumber\\
\eta'_1: & m^2(\eta'_1)= (107\,GeV)^2, \quad u_1^2=0.7333 v^2; \nonumber\\
\eta'_{2,3}: & m^2(\eta'_{2,3}) = (85.4\,GeV)^2,
\quad u_{2,3} = 0.4667 v^2; \nonumber\\
\phi_1:& mass = 100.8\, GeV; \qquad \phi_{2,3}: mass = 110.6\,GeV.
\end{eqnarray}
All numbers appear to be reasonable. Since the new objects need to be
accessed in the lepton world, it would be a challenge for our experimental
colleagues.

As for the range of validity, ${1\over 3} \le cos^2\theta_P \le {1\over 2}$.
The first limit refers to $w^2=0$ while the second for $\mu_2^2 = 0$.

We may fix up the various couplings, using our common senses. The
cross-dot products would be similar to $\kappa$, the basic coupling of
the family gauge bosons. The electroweak coupling $g$ is $0.6300$
while the strong QCD coupling $g_s=3.545$ (order of unity); my
first guess for $\kappa$ would be about $0.1$ (which is rather
small). The masses of the family gauge bosons would be estimated
by using ${1\over 2}\kappa \cdot w$, so
slightly less than $10\,GeV$. (In the numerical example with $cos
\theta_P=0.6$, we have $6 w^2= v^2$ or $w=102\,\,GeV$. This gives
$m=5\,\,GeV$ as the estimate.) So, the range of the family forces,
existing in the lepton world, would be $0.04\,\, fermi$.

In \cite{Origin}, the term that ignites the SSB is chosen to be with
$\eta'$, the purely family Higgs. This in turn ignites EW SSB
and others. It explains the origin of all the masses, in terms
of the spontaneous symmetry breaking (SSB). SSB in $\Phi(3,2)$
is driven by $\Phi(3,1)$, while SSB in $\Phi(1,2)$ from the
driven SSB by $\Phi(3,2)$, as well. The different, but
related and each self-repulsive, complex
scalar fields can accomplish so much, to our surprise. And
these Higgs are exactly those the gauge fields (i.e., the
force-fields) are calling for.

We also encounter one non-renormalization theorem of some
sort - the $\lambda$ determined in fact by the 4-dimensional
Minkowski space-time, {\it not} by the complex scalar fields
themselves. When the 4-dimensional Minkowski space-time is
given, the $\lambda$ is fixed ($\lambda={1\over 8}?$) - a
remarkable result!! How this couples with the
non-renormalization theorem and all those ultraviolet
divergences is one of the next critical questions to answer.

This "uniqueness" in the determination of the $\lambda$
means that the choice of the potential is unique \cite{Origin}
and so the Standard Model is unique. The angle $\theta_P$
is the only unknown. It also has the tremendous implication
for the ultraviolet divergences, which we proceed to discuss
in the next section.

\bigskip

\section{Maybe a way to deal wit ultraviolet divergences?}

"Is it a consistent and complete theory?" It is very difficult
question to ask, but we have to ask and try to answer. Does
the sum of all ultraviolet divergences of given order
give rise to some finite sensible number or zero? If we look at
a specific diagram, such as the self-energy diagram in Ch. 10
of \cite{Book}, ultraviolet divergence is certainly there - the
issue that troubled all famous theoretical physicists for the
entire 20th Century. Maybe in the 21st Century, there might
be some breakthroughs that would decide whether the quantum
field theory would be here to stay.

We have been rather persistent in addressing this question
- in the origin of mass \cite{Origin}, we ask this question
because we are not so sure if this solution for the origin
of mass is true or not (despite all the beautiful numbers);
in a precise definition of the Standard Model
\cite{definition}, tests on the complete theory were
discussed; and, early on, the fine-tuning problem for
introducing super-symmetry particles \cite{fine-tune}
was raised.

One general feeling that we have gained through these
exercises is that we may be doing the wrong things. The
causal $i\epsilon$ prescriptions, when combined and
manipulated further, do not always give rise to
infinities (ultraviolet divergences); rather, it
involves specification of poles and residues in the
complex energy plane. Our knowledge on a powerful
complex analysis, which still do not exist, is
rather limited. And our physics knowledge on
the complex nature of energy and of time is not
there, either.

As examples, we could try to use the procedure that
this "infinities" problem be handled by using the
dimensional regularization in the U-gauge.
Nevertheless, we are not so sure that the
answers obtained in this way would stand out.
We suspect not, sine the causal $i\epsilon$
prescriptions are ignored in the dimensional
regularization.

The other important suspicion comes from the
statement that the pure number $\lambda$
($={1\over 8}$) is the characteristic of the
4-dimensional Minkowski space-time, a number
that should be non-renormalized - the so-called
"nonrenormalization theorem". But if we
calculate the loop diagrams we will often
obtain infinities. So, we have to re-formulate
the results to make sense out of the results.

Let's assume the $3+1$ two-triplet scenario (i.e., the
standard picture) that the vacuum
expectation values (and the corresponding real
Higgs) occur for the three components of $\Phi^0(3,2)$ and the first
component of $\Phi(3,1)$ and the remaining eight make the massive
$SU_f(3)$ family gauge bosons. Note that the terms in $\delta_2$
and in $\delta_4$ vanish completely. Thus, we may draw the diagrams
for the wave-function renormalizations for the Standard-Model
Higgs, the mixed family Higgs, and purely family-triplet Higgs,
respectively, by Fig. 1, Fig. 2, and Fig. 3. Here in these figures,
we try to show only the Higgs sectors; in a complete Standard Model,
the (Dirac) fermion loop diagrams also present divergences of
quadratic order and should be dealt with simultaneously. Similarly
for the loops involving gauge bosons.

\begin{figure}[h]
\centering
\includegraphics[width=4in]{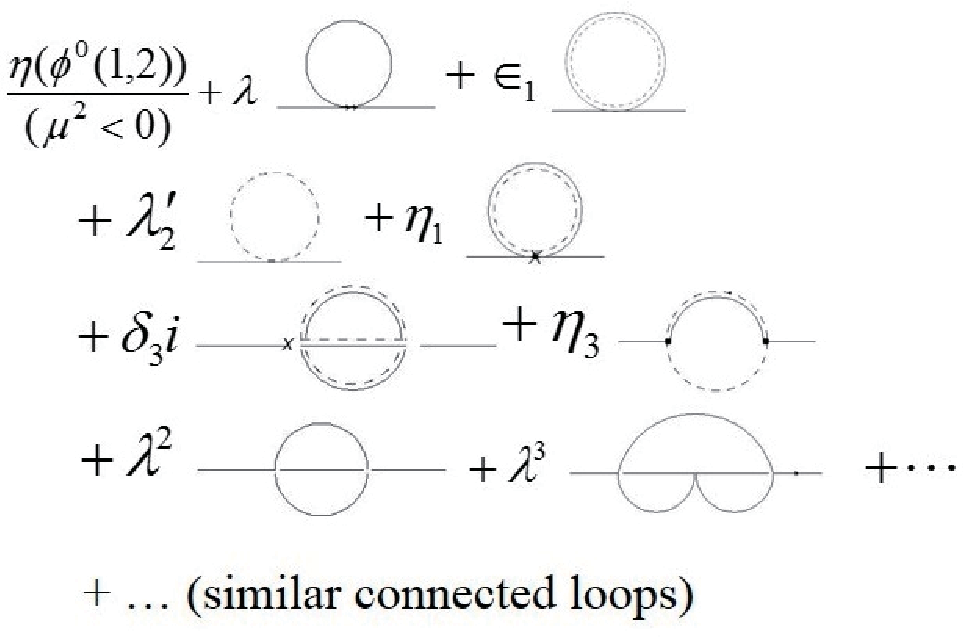}
\caption{The within-Higgs diagrams for the Standard-Model Higgs $\Phi(1,2)$.}
\end{figure}

In Fig. 1, the wave-function renormalization of the Standard-Model Higgs
$\Phi(1,2)$ is shown, for simplicity, in the U-gauge in the absence of Dirac
fermions. The lowest-order loop diagrams, from the above interaction
lagrangian, are shown from 1(b) [in $\lambda$] to 1(g) [in $\eta_3$],
where the first five are of quadratic divergence while the last one of
logarithmic divergence. The higher-order connected loop diagrams, many
of them and also of quadratic divergence, are also troublesome and should
be dealt with at the same time. We will discuss the worst divergent, i.e.,
the quadratic divergent, cases in what follows.

Using dimensional regularization (i.e. the appendix of Ch. 10, the
Wu-Hwang book, Ref. 3), we could write down the one-loop results:

We try to use one explicit example to illustrate our point related to
the infinities - the quadratic divergences of the wave function of the
SM Higgs $\eta$. We may draw the diagrams
for the wave-function renormalizations for the Standard-Model
Higgs, the mixed family Higgs, and purely family-triplet Higgs,
respectively. The result for the Standard-Model Higgs is shown by
Fig. 1. Here, in Fig. 1, we try to show only the Higgs sectors
themselves; in a complete Standard Model, the (Dirac) fermion
loop diagrams, and those with gauge bosons, also present divergences
of quadratic order and should be dealt with simultaneously.

We don't know if the procedure is right or not. But it is interesting
to know that the formulae in dimensional regularization work and that
in fact it works in the U-gauge. For example, the $Z^0-boson$ loop
for Fig. 1 would give us the vanishing result - so, it does not
bother us.

In details, the coupling of the SM Higgs is (\cite{Book}, e.g.,
Wu/Hwang, Ch. 13)
\begin{equation}
-{1\over 8}(v^2+2v\eta +\eta^2) \{2g^2 W_\mu^+W_\mu^-
+ [g^2+(g')^2] Z_\mu^0 Z_\mu^0\},
\end{equation}
which gives rise to, to the first order, the one-loop $W^\pm$
or $Z^0$ diagram. To evaluate them, we use the propagator
in the U-gauge (see the appendix of Ch. 13) and the formulae
in the dimensional regularization (see the appendix of Ch. 10).
They cancel between two terms for each diagram.

To proceed, we examine those diagrams in Fig. 1 which are
"simple" quadratically divergent - those at the one-loop order.
These are among the various Higgs.

In Fig. 1, we show the wave-function renormalization of the Standard-Model Higgs
$\Phi(1,2)$, among the Higgs, in the U-gauge. The lowest-order loop diagrams,
from the above interaction lagrangian, are shown from 1(b) [in $\lambda$]
to 1(g) [in $\eta_3$], where the first five are of quadratic divergence while
the last one of logarithmic divergence. The higher-order connected loop
diagrams, many of them and of quadratic divergence multiplied by logarithmic
divergences, are also troublesome.

The one-loop diagrams involving the quark (or charged lepton),
when simplified, are sums of quadratic and logarithmic
divergences.

Using dimensional regularization (i.e. the appendix of Ch. 10, the
Wu-Hwang book, Ref. \cite{Book}), we obtain the one-loop and
quadratic-divergence results as follows. In the dimensional
regularization, the factor $\Gamma(1-{n\over 2})$ stands for where
the quadratic divergence appears. Maybe the fractional dimensions,
which are represented as finite numbers, could get some meaning,
but we have to remember that, as a drawback, we bypass the
$-i\epsilon$ in the propagators.

In follows, we concentrate only on those ultraviolet
divergences of quadratic order:

\begin{eqnarray}
& -4\cdot {n\over 2}\cdot (S_q + S_{c.l.})\Gamma(1-{n\over 2}) \nonumber\\
&+ \{ 3\lambda m^2(\eta) + {\epsilon_1\over 2} \sum_i m^2(\eta'_i)
+ \epsilon_1 \sum_i m^2(\phi_i)\nonumber\\
& +{\lambda_2^\prime\over 2} m^2(\eta') +
{\eta_1\over 2} \sum_i m^2(\eta'_i)\}\Gamma(1-{n\over 2})  \sim 0;\\
& S_q = \sum_{quarks} 3\cdot G_i^2\cdot (m_i^2 -{1\over 6} m^2(\eta)),\nonumber \\
& S_{c.l.} = \sum_{c.l.} G_i^2 \cdot (m_i^2 -{1\over 6} m^2(\eta)).
\end{eqnarray}
Or, using the Standard Model, we have
\begin{eqnarray}
& -4\cdot {n\over 2}\cdot (S_q+S_{c.l.})\Gamma(1-{n\over 2})\nonumber\\
& + \{\lambda (3m^2 (\eta) - cos\theta_P\sum_i m^2(\eta'_i) + 2 cos\theta_P
\sum_i m^2(\phi_i)) + {\lambda'_2\over 2} m^2(\eta')\} \Gamma(1-{n\over 2}) \sim 0.
\end{eqnarray}
Here we should give a few words about "$\sim 0$". Since in the
dimensional regularization one ignores the causal $i\epsilon$
prescriptions altogether, the presumed "cancelations" also may
be lost.

Basically, we first focus our attention only on the quadratic
divergences, since these are "the highest divergences" in the
lowest loops - relatively easy to "collect" and the most
important divergences altogether; if we cannot do
anything about them, then the game is over.

Here the coefficients of $\Gamma(1-{n\over 2})$ are the coefficients of
quadratic divergences while those of $\Gamma(2-{n\over 2})$ are the
coefficients of logarithmic divergences - for the latter, divergence is
less severe and the contributions could be everywhere. The dimensional
regularization does help us to get something useful, if they are
useful.

\begin{figure}[h]
\centering
\includegraphics[width=4in]{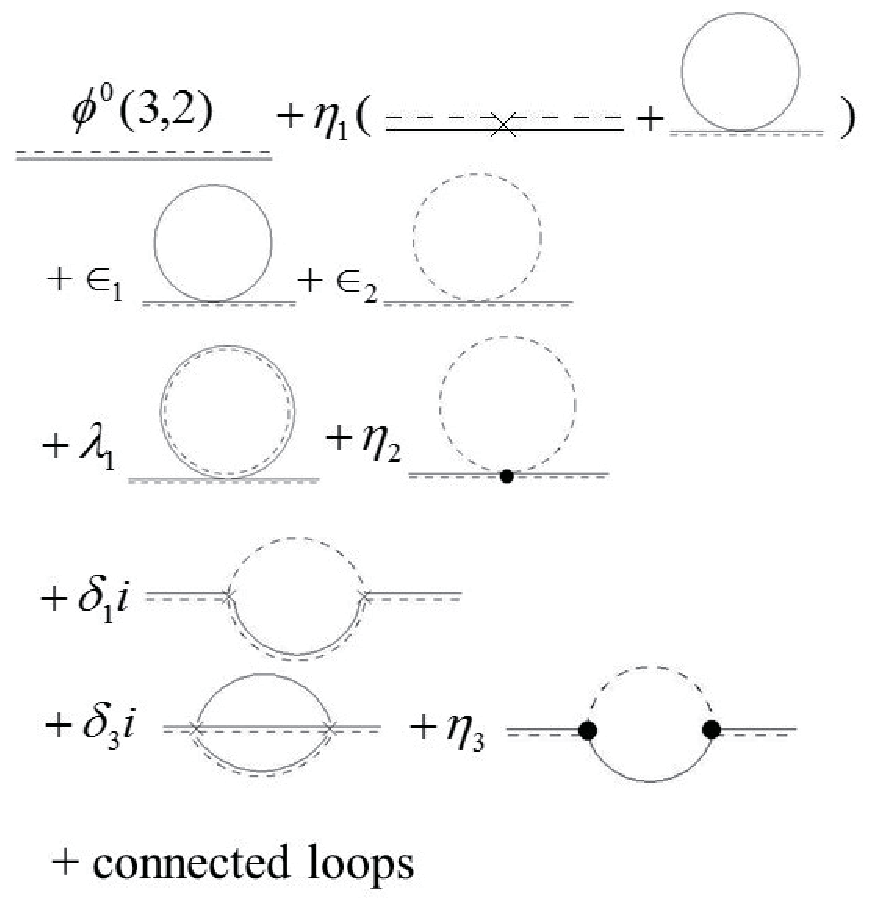}
\caption{The diagrams for the mixed family Higgs $\Phi(3,2)$.}
\end{figure}

\begin{figure}[h]
\centering
\includegraphics[width=4in]{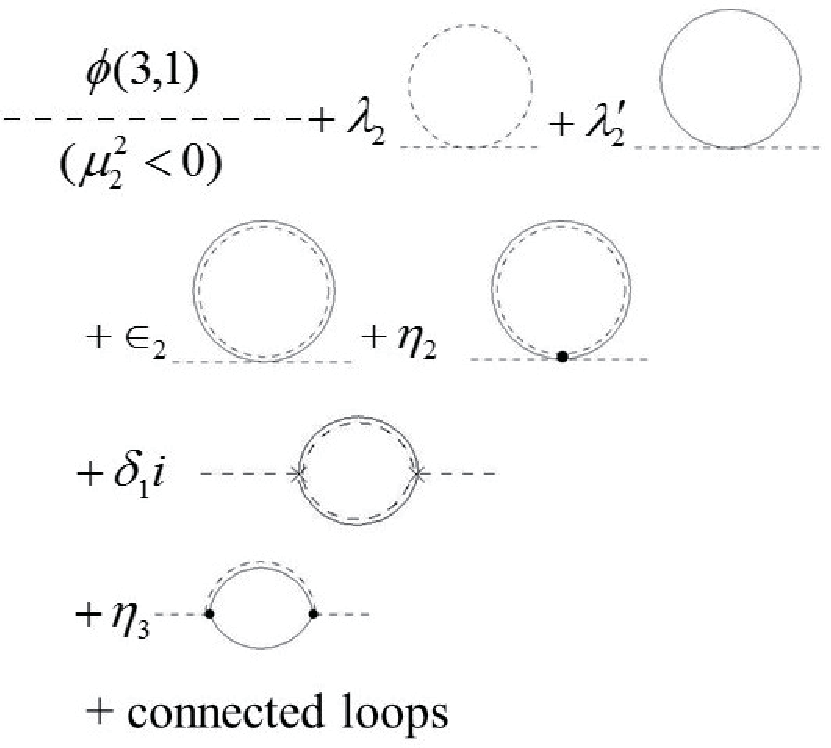}
\caption{The diagrams for the purely family Higgs $\Phi(3,1)$.}
\end{figure}

Figs. 2 and 3 show the wave-function renormalizations of family Higgs
$\Phi(3,2)$ and $\Phi(3,1)$ with pretty much the same story. They are
coupled altogether. Here the quarks do not enter at all but the off-diagonal
elements in the lepton world might cause some problems.

On Figs. 2, we have for $\eta'_1$, again for quadratic divergences,
\begin{eqnarray}
& -4\cdot {n\over 2}\cdot T_{lepton}\Gamma(1-{n\over 2}) \nonumber\\
&+ \{ 3\lambda_1 m^2(\eta'_1) + ... (mixed) +
{\epsilon_1+ \eta_1\over 2} m^2(\eta) \nonumber\\
& +
{\epsilon_2+\eta_2\over 2} m^2(\eta')\}\Gamma(1-{n\over 2})  \sim 0;\\
& T_{lepton} = \sum H_i^2 \cdot (m_i^2 -{1\over 6} m^2(\eta'_1)).\nonumber
\end{eqnarray}
Here $T_{lepton}$ is defined in accordance with the curl-dot
product in neutrinos or in charged leptons. Since the sign
switch in $\eta_1$ or in $\eta_2$, the overall cancelation
is always possible.

For the family Higgs $\eta_2$ and $\eta_3$ it is easy to duplicate
the results. Note that the charged scalar fields, i.e. the charged
part of $\Phi(3,2)$, do play some roles in this game.

On Figs. 3, we have for $\eta'$, also for quadratic divergences,
\begin{eqnarray}
& \{ 3\lambda m^2(\eta') + {\epsilon_2+\eta_2\over 2} \sum_i m^2(\eta'_i)
+ \epsilon_2 \sum_i m^2(\phi_i)\}\Gamma(1-{n\over 2})  \sim 0.\nonumber\\
\end{eqnarray}
We note that the term in $\eta_2$ could cancel the other two - in hoping
to make complete the cancelation.

Here in the last two equations we need to examine whether the loops in
the family gauge bosons would make the contributions for the problem
of quadratic divergences - for the sake of the overall cancelation.
Again, we note the meaning that should be attached to $\sim 0$.

The diagrams which are of quadratic divergence are troublesome
since, as shown in Fig. 4, the series could be blown-up, of
$2n$-th divergence with $n \to \infty$. Mathematically, we should
avoid such terms by all means. Well, some of these "strange"
things occur in higher orders in the dimensional
regularization.

We are hinting also that we have to study the mathematics of
"divergences"; maybe they are there, because of the
uncountably infinite degrees of freedom and other reasons,
and there are regularities to be discovered \cite{fine-tune}.

\begin{figure}[h]
\centering
\includegraphics[width=4in]{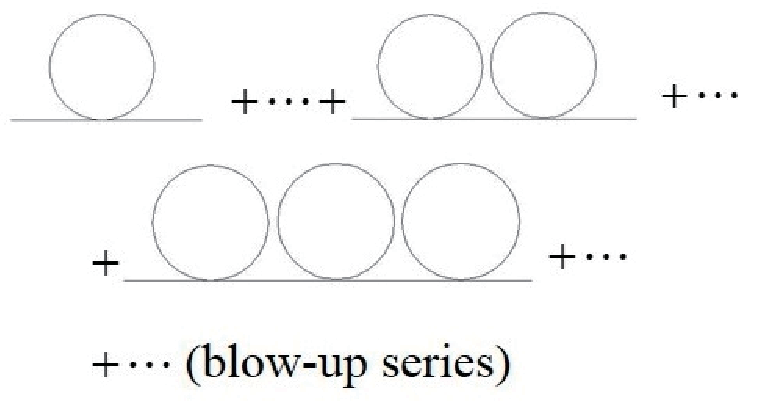}
\caption{The blow-up diagrams which are repetitive in quadratic divergences.}
\end{figure}

To the one-loop order, Fig. 1, plus fermion one-loops, gives
rise to the results for quadratic divergences. As pointed out earlier,
numerous two loops and high-order loops also give rise to quadratic
divergences (and logarithmic divergences as the residual). In
perturbation series, $\lambda_i^n$, $(n = 2 \,\,or\,\, larger)$,
would make finite terms smaller, or even negligible, but the
divergent terms infinite. We are not so sure that the conjecture
of cancelation \cite{fine-tune} can be proved - all quadratic
divergences cancel out (noting that this is more important) and
logarithmic divergences also cancel out (meaning that we are back
in treating a finite theory). Note that we do not believe that
the dimensional regularization is the way to go, as far as
the cancelation theorem is concerned.

In fact, we notice further that, according to dimensional-regularization
results, the three-loop diagram gives rise to (quadratic
divergence) $\times$ (logarithmic divergence), and so on. They
have to be organized differently. One simple
way out is that they cancel completely in their own group, such as
all the four-loop diagrams.

To summarize from Figs. 1-4 and the corresponding results
from the dimensional regularization, we do obtain some
useful results, judging from the extremely difficult
tasks associated with "ultraviolet divergences". But this
is obviously not the whole story, since, as said earlier,
the dimensional regularization
does not give the causal results - the so-called
$i\epsilon$ results.

In the above, the cancelations for the results indicated
by "$\sim 0$" are in fact not there. We go in length
about the "wrong" (or, maybe wrong,) infinite results,
mainly because we need to eventually find a better
(right) way. In search of the "right" physics, we have
to put everything together and try to sort out of the
right part of the story.

In fact, the $i\epsilon$ causal prescriptions specify the
poles in complex analysis - if we are honest about them,
some of infinities might not be "infinities" any longer.
In the dimensional regularization, we ignore the causal
$i\epsilon$ prescriptions altogether; we might have to
come back to Pauli-Villars regularization or others
\cite{Book} but select more physically what needs to be
subtracted (and to be understood in the physics-wise
sense). In the 21st Century, this should be one of
the main tough tasks in quantum field theory.

Maybe at this juncture we should speculate a little
more on this causal $i\epsilon$ prescription. In the
simplest cases, it gives us the propagators. Everything
should be in the sense of complex analysis (mathematics)
and, to deal with, there are integrals of products of
propagators. Beside Cauchy's theorem, we need many more
comprehensive theorems in complex analysis to play with.
(We still do not have any of these theorems, yet.)
Physics-wise, our endless attempts seem to extend the
energy, or the time, to become a complex variable.
Maybe what we try to do is not something infinite -
rather an illusion of the complex variables viewed
as some real energy, or as some real time.

So, the dimensional regularization, Pauli-Villars, etc.
\cite{Book}, efforts of the 20th Century might have
missed the point - the time or the energy of our
world is complex, truncated as though it is real
to us. Hopefully, we could begin the 21st Century
with the eyes of completely new visions - after all,
we declare that we live in the quantum 4-dimensional
Minkowski space-time with the force-fields $SU_c(3)
\times SU_L(2) \times U(1) \times SU_f(3)$ gauge-group
structure built-in from the very beginning.

\bigskip

\section{Some Insights in Physics}

One way to verify the Standard Model is the experimental search
for the family Higgs $\eta'_1$, or $\eta'_{2,3}$, or charged family
Higgs $\phi_1^+$ and $\phi^+_{2,3}$, or pure family
Higgs $\eta'$, in a $200\,GeV$ $e^-e^+$ collider, since these family
particles can only be accessed in the lepton channels. Maybe it was
a little early to shut down the LEP-II operations at CERN.

The active search is through "the family collider \cite{collider}",
of a $\mu^\mp e^\pm$ collider, since two generations of leptons must
be simultaneously involved in the search of family Higgs $\eta'_1$.
The technology might be not quite ready in developing the
"unstable" $\mu^\pm$ beams; thus, the option of the $e^-e^+$
collider in this study should be there.

The implication of the family gauge theory is in fact a multi-GeV or
sub-sub-fermi gauge theory - the leptons are shielded from this
$SU_f(3)$ theory against the QED Landau's ghost. An active
search of this force clearly should be encouraged. The $g-2$ anomaly
should certainly deserve another serious look in this context.

In all the Standard Model, the GIM mechanism in fact makes
the masses of down-type quarks off-diagonal, even though quarks
are the singlets in the $SU_f(3)$ space. Under $SU_f(3)$, the masses
of the three charged leptons are $m_0 + a \lambda_2 + b \lambda_5 +
c \lambda_7$ (before diagonalization) while the masses of neutrinos are
purely off-diagonal, i.e. $a' \lambda_2 + b' \lambda_5 + c' \lambda_7$.
This result is very interesting and very intriguing. How to develop
a formalism with the off-diagonal masses should be the important
task of all the theoretical physicists \cite{TMYan}.

This result follows from the above curl-dot product, or, the $\epsilon^{abc}
\bar \Psi_{L,a} \Psi_{R.b} \Phi_c$ product, i.e. the
$SU_f(3)$ operation, in writing the coupling(s) to the right-handed
lepton triplets. In fact, we have $a'/a^*=b'/b^*=c'/c^*$ for the coupling
strengths. QCD is also $SU(3)$ and baryons are constructed of three
triplets of quarks - our studies of $SU(3)$ could go deeper yet.

In addition, neutrinos oscillate among themselves, giving rise to a
lepton-flavor-violating interaction (LFV). There are other
oscillation stories, such as the oscillation
in the $K^0-{\bar K}^0$ system, but there is a fundamental "intrinsic"
difference here - the $K^0-{\bar K}^0$ system is composite while neutrinos
are "point-like" Dirac particles. We have standard Feymann diagrams for
the kaon oscillations but similar diagrams do not exist for point-like neutrino
oscillations - our proposal solves the problem, maybe in a unique way.

Thinking it through, it is true that neutrino masses and neutrino
oscillations may be regarded as one of the most important experimental
facts over the last thirty years \cite{PDG}. Treating neutrinos as
"point-like" Dirac particles, neutrinos oscillations between different
generations indeed present us some fundamental questions.

In fact, certain LFV processes such as $\mu \to e + \gamma$ \cite{PDG},
$\mu + A \to A^* + e$, $e^+ + e^- \to \mu^+ + e^-$, etc., are closely
related to the most cited picture of neutrino oscillations \cite{PDG}. In
recent publications \cite{Hwang10}, it was pointed out that the cross-generation
or off-diagonal neutrino-Higgs interaction may serve as the detailed mechanism
of neutrino oscillations, with some vacuum expectation values of the family Higgs,
$\Phi(3,1)$ and $\Phi^0(3,2)$. So, even though we haven't seen, directly,
the family gauge bosons and family Higgs particles, we already see the
manifestations of their vacuum expectation values.

\bigskip

\section{Closing Remarks}

{\it Everything associated with our Standard Model, except the
"ignition" term, is dimensionless - all couplings with the force
fields (gauge fields), the complex scalar Higgs fields, the quark
world, and the lepton world, all of them are dimensionless. Hence,
it is determined by the quantum 4-dimensional Minkowski
space-time.}

To close this paper, we would like to comment on the "minimum Higgs
hypothesis" and "Dirac similarity principle".

In a slightly different context \cite{Hwang3}, it was proposed earlier
(five years ago) that we could work with two working rules: "Dirac
similarity principle", based on eighty years of experience, and
"minimum Higgs hypothesis", from the last forty years of experience.
Using these two working rules, the extended model mentioned above
becomes rather unique - so, it is so much easier to check it
against the experiments. To move forward in building up our
knowledge, there are moments that we have to play conservatively
- at this moment, we introduce the so-called rules.

We have to say that the phenomenon of three generations is one leading
puzzle in particle physics; nowadays, it is safe to add that neutrino
oscillations is another puzzle. To understand these puzzles, we have
to admit that the natural consequence is the $SU_f(3)$ family gauge
theory - to put everything together, it leads to our Standard
Model \cite{Hwang417}, with a unique "generalized" joint Higgs
mechanism. These Higgs, if alone, will be self-repulsive
and, if jointly, the mutual attractive forces will keep them
there - thus they exist. It is in accord with the "minimum
Higgs hypothesis".

Now, we understand \cite{definition} that our Space is the
$SU_c(3) \times SU_L(2) \times U(1) \times SU_f(3)$ Minkowski
space-time, that can only accommodate the scalar fields $\phi$
with a natural-born $\lambda (\phi^\dagger \phi)^2$ "repulsive"
self-interaction only in the exceptional cases (when they
could become the longitudinal components of the gauge field).
This explains "minimum Higgs hypothesis".

For "Dirac similarity principle", it seems rather powerful
that the Dirac linearization of the Einstein's relation,
$E^2={\vec p\,}^2 + m^2$, leads to the concept of the spin
and also the concept of the antiparticle. To us, when
they are consistent with (123) or another (123), the
"background" of our world accepts them (the quark world or
the lepton world) for free (without costing the
energies). It goes with the quark world and also with the
lepton world. Everything is so perfect, both physically and
mathematically.

We may add that, under two working hypotheses, we can close the
Universe; that is, all the dark-matter particles and all the
ordinary-matter particles are accounted for. Our Standard Model
provides a description of the entire world - the 25\% dark-matter
world and the 5\% ordinary-matter world.

\end{document}